\documentclass[%
 reprint,
superscriptaddress,
nofootinbib,
 amsmath,amssymb,
 aps, pra,
floatfix,
]{revtex4-2}

\usepackage{graphicx}
\usepackage{dcolumn}
\usepackage{bm}
\usepackage {xcolor}

\usepackage{physics}
\usepackage{hyperref}
\hypersetup{
    pdftitle={OptimalPOVMTomography},
    pdfauthor={BacuiLi},
    pdfsubject={OptimalPOVMTomography},
    bookmarksnumbered=true,     
    bookmarksopen=true,         
    bookmarksopenlevel=1,       
    colorlinks=true,            
    pdfstartview=Fit,           
    pdfpagemode=UseOutlines, 
    pdfpagelayout=TwoPageRight,
    breaklinks=true}
\usepackage{hypcap}
\usepackage{soul}
\usepackage{xurl}
\hypersetup{breaklinks=true}
\def\ANU{Centre for Quantum Computation and Communication Technology, Department of Quantum Science, Australian National University, Canberra, ACT 2601, Australia.}
\def\Astar{Institute of Materials Research and Engineering, Agency for Science, Technology and Research (A*STAR), 2 Fusionopolis Way, 08-03 Innovis 138634, Singapore}
\begin{document}
\preprint{APS/123-QED}

\title{Optimal Single Qubit Tomography:\\Realization of Locally Optimal Measurements on a Quantum Computer}

\author{Bacui Li}
\email{patrick.li.bj@qq.com}
 \affiliation{\ANU}
\author{Lorc\'{a}n O. Conlon}
\email{lorcanconlon@gmail.com}
 \affiliation{\ANU}
\author{Ping Koy Lam}
\affiliation{\ANU}
\affiliation{\Astar}
\author{Syed M. Assad}
\email{cqtsma@gmail.com}
\affiliation{\ANU}

\date{\today}

\begin{abstract}
Quantum bits, or qubits, are the fundamental building blocks of present quantum computers. Hence, it is important to be able to characterize the state of a qubit as accurately as possible. By evaluating the qubit characterization problem from the viewpoint of quantum metrology, we are able to find optimal measurements under the assumption of good prior knowledge. We implement these measurements on a superconducting quantum computer. Our demonstration produces sufficiently low error to allow the saturation of the theoretical limits, given by the Nagaoka--Hayashi bound. We also present simulations of adaptive measurement schemes utilizing the proposed method. The results of the simulations show the robustness of the method in characterizing arbitrary qubit states with different amounts of prior knowledge. 
\end{abstract}

\maketitle

\section{Introduction}
The qubit is the basic unit of information in most quantum computers. Qubit-based quantum computers have shown great promise in sensing~\cite{Marciniak2022OptimalSensors} and computation technologies~\cite{WuAdvantage2,Arute2}. Thus, it is essential for the development of quantum computers to study qubit states in detail. The field of research which focuses on characterizing quantum states is called state tomography~\cite{Nielsen2010QuantumEdition,Rehacek2004MinimalTomography,Wootters1989OptimalMeasurements}. More specifically, state tomography concerns the problem of finding the most accurate method of reconstructing the state with a fixed amount of quantum resources, or equivalently, finding the most efficient method that reconstructs the state up to some fixed threshold uncertainty with the minimal amount of quantum resources.

Any reconstruction method will involve measurements of the quantum state. All physically possible measurements in the context of quantum systems are described by a positive operator-valued measure (POVM). Among all possible POVMs, projective measurements form a special subset. Projective measurements can be described by a single observable operator and are orthogonal to one another. To reconstruct a quantum state without ambiguity, we require that the measurement used is informationally complete~\cite{Czerwinski2021QuantumDomain}. With certain measurements, the information obtained is insufficient to reconstruct the state. For example, a projective measurement along the $z$-axis of a single qubit state cannot provide any information about the $x$ or $y$ components of the Bloch vector and this measurement alone will fail to reconstruct an arbitrary state.

Two reconstruction methods that are widely regarded as optimal are projective measurements onto mutually unbiased bases (MUB)~\cite{Wootters1989OptimalMeasurements} and the symmetric informationally complete POVM (SIC-POVM)\cite{Zhu2010SICDimensions,Renes2004SymmetricMeasurements}. Both of these measurements have been realized experimentally \cite{Li2015RealizationExperiment,Zhao2015ExperimentalWalks,Bian2015RealizationWalk} and the SIC-POVM has already been demonstrated for quantum entanglement detection \cite{Bae2019Linking2-designs,Li2021EntanglementMeasurement} and tomography protocols \cite{Zhao2015ExperimentalWalks,Medendorp2011ExperimentalMeasurementsb,Bian2015RealizationWalk,Hou2018DeterministicWalks}. However, for reconstructing an unknown state, the repeated measurement of an ensemble of identical unknown states with either the SIC-POVM or MUB will not be the optimal method. This is because neither of these measurements exploits the knowledge gained from previous measurement results. Indeed, the optimality of the SIC-POVM for state reconstruction has only been shown under the condition of measuring a completely unknown state uniformly distributed following the Haar measure of the unitary operators \cite{Scott2006TightMeasurements,Petz2012EfficientMeasurements}. Thus, when we have partial knowledge of the state in question, other measurements may become optimal.

The optimality of the SIC-POVM and MUB is questioned in a localized setting, where the quantum state is well known and we attempt to sense small changes in the state. This bears great similarity to the setting used in quantum metrology, where physical parameters are encoded onto quantum states, and we attempt to sense small changes in those parameters \cite{Holevo2011ProbabilisticTheory}. It is natural, therefore, to utilise tools from quantum metrology to analyse quantum state tomography in the localized setting. In quantum metrology, it is common to use Cram\'{e}r--Rao bounds to place limits on how accurately physical quantities can be measured \cite{Holevo2011ProbabilisticTheory,Helstrom1967MinimumStatistics,Helstrom1974NoncommutingTheory}. One bound which is of particular practical relevance is the Nagaoka--Hayashi (N-H) bound, as it applies when probe states are measured individually \cite{Nagaoka2005AEstimation,Conlon2021EfficientMeasurements}. In this work, we map the qubit tomography problem to quantum metrology, allowing us to use the N-H bound as an experimentally accessible bound for tomography. The N-H bound, thus, provides a reference against which the measurement proposed in this paper can be compared.

Under the local setting, a measurement inspired by MUB has been proposed. This measurement used a probabilistic mixture of three projective measurements that are orthogonal to each other, such as projective measurements on the $x,y$ and $z$ axes of the Bloch sphere and was shown to be optimal for the single-qubit state reconstruction problem \cite{Hou2016AchievingTomography}. In this work, we investigate an optimal reconstruction using a POVM inspired by the SIC-POVM. The SIC-POVM can be represented by an equiangular constellation of complex vectors in Hilbert space. For qubit states, the SIC-POVM, therefore, forms a tetrahedron. The POVM we propose is similar, being a tetrahedron squashed along one axis. Like the SIC-POVM, our measurement requires only 4 POVM elements, the minimal number for an informationally complete measurement on a single qubit state. 

From a theoretical perspective, the connection between our proposed POVM and the SIC-POVM suggests the existence of some unexplored optimal POVMs in higher dimensions\footnote{Although the existence of such an object for arbitrary dimension $N$ is still an open problem \cite{Horodecki2022FiveTheoryb}} given any specific amount of prior knowledge. From a practical viewpoint, we anticipate that the proposal and implementation of an optimal measurement for localized state tomography will stimulate the development of optimal adaptive protocols for quantum tomography that converge to the locally optimal measurement we proposed.


The structure of this paper is as follows: in Section.\;\ref{sec:bg}, we give a detailed description of the state tomography problem in the localized setting. We demonstrate that the SIC-POVM, which is optimal for uniformly distributed states, does not saturate the \mbox{N-H} bound. Hence, it is not optimal in this context. In Section.\;\ref{sec:OptPOVM}, we introduce optimal POVMs that saturate the \mbox{N-H} bound. We present the analytical form of the POVMs, which depend on the density matrix of the mixed state $\rho$. 

In Section.\;\ref{sec:QCMeth}, we show the experimental procedure that implements the optimal POVM on a quantum computer. The method proposed is robust for estimating all mixed states except for marginal cases when the purity is exactly 1. Here we utilise an efficient circuit decomposition of a two-qubit unitary~\cite{Vatan2004OptimalGates}, which uses fewer CNOT gates and single-qubit gates than the previous approach of implementing qubit POVMs~\cite{Yordanov2019ImplementationComputer}. The processing of the raw quantum circuit output integrates measurement gate error mitigation \cite{Bravyi2020MitigatingExperiments} and a live calibration scheme which ensures the unbiasedness of the measurement and improves the overall performance. These error mitigation approaches have been shown to improve the quality of results in other areas of quantum information research \cite{Yamamoto2021Error-mitigatedPurification,Mooney2021GenerationComputer,Maciejewski2019MitigationTomography}. 

In Section.\;\ref{sec:QCexp}, we show the results generated by methods discussed in Section.\;\ref{sec:QCMeth}. The results validate the feasibility and demonstrate the optimality of the proposed POVMs. Nonetheless, due to various noises in the current generation of quantum computers, the resulting mean squared error (MSE) value has room for improvement, as the resulting MSE is at worst 15\% above the predicted theoretical MSE.

Finally, in Section.\ref{sec:Comp_Sim}, we will discuss the application of the proposed POVMs in a broader context and address the limitation of the method when purity is close to or is exactly 1. We address one of the major drawbacks of the localized approach, namely that a good prior knowledge of the state is required. We present a two-step adaptive method incorporating both the SIC-POVM and our proposed POVMs. This adaptive method outperforms either the SIC-POVM or the proposed POVM when either of them is used independently. We also consider a generalisation of our optimal POVM for arbitrary amounts of prior knowledge. In the limit of a completely unknown state, this reduces to the SIC-POVM as expected. The result from simulations of these two adaptive measurement approaches demonstrates the robustness of the method without the assumption of good prior knowledge.

\section{Background \label{sec:bg}}
To find the optimal measurement in the local setting discussed above, we first define the problem and the figure of merit, which will be used throughout the paper. The tomographical problem we will investigate is formulated as a linear quantum state tomography problem \cite{Petz2014ConditionalSIC-POVMs}: We would like to reconstruct an arbitrary mixed single-qubit state linearly parameterised as
\begin{equation}
    \rho = (I + \Vec{\sigma}\cdot\boldsymbol{\theta})/2\;,
    \label{eq:rho}
\end{equation}
where $\Vec{\sigma}$ is the vector of Pauli matrices and $\boldsymbol{\theta}$ is the vector of parameters we wish to learn about. We define $r$ as the length of the vector $\boldsymbol{\theta}$, i.e., $r = \abs{\boldsymbol{\theta}}$. $r$ also determines the purity by $(r^2+1)/2$ and is a crucial parameter for the qubit state as we shall see in the later sections.
For the estimated state, $\hat{\rho}$, the metric of deviation we use is the trace norm $D_{\rho,\hat{\rho}} = \textrm{Tr}[\sqrt{(\hat{\rho}-\rho)^\dagger(\hat{\rho}-\rho)}]$ (twice the trace distance). In this formulation, the trace norm equals the Euclidean distance between the estimated parameter vector $\hat{\boldsymbol{\theta}}$ and the true parameter vector $\boldsymbol{\theta}$, i.e., $D_{\rho,\hat{\rho}} = |\hat{\boldsymbol{\theta}}-\boldsymbol{\theta}|$ where $\hat{\boldsymbol{\theta}}$ is the estimated vector of parameters $\boldsymbol{\theta}$ \cite{Holevo2011ProbabilisticTheory}. 

This formulation enables us to minimize the expectation value of the square of the trace norm between the true state and the estimated state, $E[D_{\rho,\hat{\rho}}^2]$ by minimizing the total MSE of the three parameters. In addition to the formulation presented above, some studies also consider the Hilbert--Schmidt norm or fidelity as their figure of merit \cite{Hou2018DeterministicWalks,Scott2006TightMeasurements}. In the case of qubit state tomography, the Hilbert--Schmidt norm is proportional to the trace norm between two arbitrary mixed states, thus, it is equivalent to minimizing either of the two norms. Under the formulation discussed above, the MSE produced by the SIC-POVM for a mixed state is \cite{Rehacek2004MinimalTomography}:
\begin{equation}
\label{eq:MSE_SIC}
    \textrm{MSE}_{\textrm{SIC}} = 9-r^2\;.
\end{equation}

While the MSE that the SIC-POVM can attain is optimal in the non-local setting, the development of quantum metrology theory suggests that better measurements may be possible in the local setting we are considering. One such development is the precision bound for multiparameter estimation, the N-H bound \cite{Nagaoka2005AEstimation,Conlon2021EfficientMeasurements}. This bound generalizes information trade-off relations such as the Heisenberg uncertainty principle and gives the precision limit of any multiparameter estimation problem when the probe states are measured individually. The MSE described above can be lower bounded by the N-H bound as we have essentially transformed the problem of linear quantum state tomography into a multiparameter estimation problem \cite{Nagaoka2005AEstimation}, for which the N-H bound is applicable. Facilitated by the method of semi-definite programming for efficient calculation of the N-H bound \cite{Conlon2021EfficientMeasurements}, we verified that the N-H bound for the single qubit tomography problem is
\begin{equation}
    C_{\textrm{NH}} = \left(2+\sqrt{1-r^2}\right)^2\;.
\end{equation}
This shows that the MSE for estimating $\boldsymbol{\theta}$ obtained from identical measurements on $N$ probe states, MSE$(N)$ when measuring them one by one is lower bounded by $C_{\textrm{NH}}/N$ for all unbiased measurement methods\footnote{This bound agrees with the bound derived by R. D. Gill and S. Massar \cite{Gill2000StateEnsembles,Hou2016AchievingTomography}.}. A measurement that reaches the value of $C_{\textrm{NH}}/N$ will be deemed optimal. It is easy to show that $C_{\textrm{NH}}<\textrm{MSE}_{\textrm{SIC}}$ except at $r=0$. Thus, the SIC-POVM is not optimal, and there may exist an optimal measurement that performs better than the SIC-POVM and saturates the N-H bound.

\section{Optimal POVM\label{sec:OptPOVM}}

\begin{figure}[b]
\includegraphics[width = \linewidth]{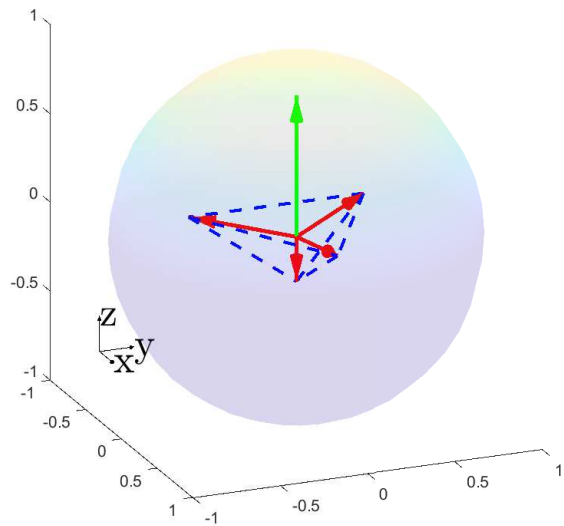}
\caption{\label{fig: STH_vis} A Bloch sphere representation of the four elements of the ST-POVM (with $r_p = 0.8$ and $\phi = 0$) shown by the red arrows. The green arrow shows the state being estimated. The shape formed by the four vertices is a triangular pyramid. When the state being estimated lies closer to the surface of the Bloch sphere (higher purity), the shape formed by the 4 elements is squashed further. When estimating the maximally mixed state, $r_p$ goes to zero, and the ST-POVM converges to the SIC-POVM.}
\end{figure}

We now present the analytical form of the optimal POVMs which saturate the N-H bound. Depending on the mixed state, $\rho$ (as parameterised in Eq.\;\ref{eq:rho}), we have different solutions for the optimal POVM, $\{\Pi_i(\boldsymbol{\theta})\}$. Each POVM consists of 4 rank-1 operators. There is a simple decomposition for the elements, $\Pi_i = r_i\ket{\psi_i}\bra{\psi_i}$ where $r_i$ is the real amplitude and $\ket{\psi_i}$ is a pure-state state vector. Notice that the amplitude $r_i$ denotes the trace of the corresponding POVM element, i.e., $\textrm{tr}(\Pi_i) = r_i$. Hence, we can visualize the POVM element $\Pi_i$ by visualizing a vector in the Bloch sphere whose direction is defined by the direction of $\ket{\psi_i}$ and its length by $r_i$. The four vectors are orientated such that the representation of the POVM element, $\Pi_z$, points along the opposite direction of the probe state $\rho$ while the other 3 points to vertices of an equilateral triangle in the opposite direction. A POVM in this symmetrical shape is inspired by the SIC-POVM which forms a tetrahedron, thus we call the novel POVM the \textit{Squashed Tetrahedron} POVM (ST-POVM). 

The ST-POVM is parameterized by two parameters, $0\leq r_p <1$ which determines the ``squashedness'' of the tetrahedron and the angle $\phi$ which orientates the legs forming the equilateral triangle. Explicitly, the four POVM outcomes are given by $r_z = \left(1+\sqrt{\frac{1+r_p}{1-r_p}}\right)^{-1}$, $r_{1} = r_{2} = r_{3} = \frac{2-r_z}{3}$ and 
\begin{equation}
\begin{aligned}
    \ket{\psi_z} &= \ket{1}\\
    \ket{\psi_1} &= \ket{0}A_0 + \ket{1}A_1 e^{i\phi}\\
    \ket{\psi_2} &= \ket{0}A_0 + \ket{1}A_1 e^{i(\phi+2\pi/3)}\\
    \ket{\psi_3} &= \ket{0}A_0 + \ket{1}A_1 e^{i(\phi+4\pi/3)}\;,
\end{aligned}
\end{equation}
where $A_0=1/\sqrt{3r_1}$ and $A_1 = \sqrt{1-1/3r_1}$. These 4 POVM elements with $\phi = 0$ are shown in Fig.\;\ref{fig: STH_vis}.

With this POVM and setting $\phi = 0$, we obtain an estimate of $\boldsymbol{\theta}$ from $\hat{\theta}_j = \sum_k \mathcal{E}_{jk} f_k$, where
\begin{equation}
\label{eq:EstimateTheta}
\mathcal{E} = 
\begin{pmatrix}
0 & 2a & -a & -a\\
0 & 0  & \sqrt{3} a & -\sqrt{3} a\\
b & 1 & 1 & 1
\end{pmatrix}\;,
\end{equation}
$a = \sqrt{1+\sqrt{\frac{1-r_p}{1+r_p}}}$ and $b = -1-2\sqrt{\frac{1+r_p}{1-r_p}}$. $f_k = n_k/N$ $(k = z,1,2,3)$ are the frequencies with which each POVM outcome occurs given $N$ probes are measured and the outcome $\Pi_k$ was observed $n_k$ times. One can verify the unbiasedness of $\hat{\boldsymbol{\theta}}$ for all $r_p$ and $\phi$ by calculating $\mathbb{E}[\hat{\theta}_j]$. This is given by $\mathbb{E}[\hat{\theta}_j] = \sum_k \mathcal{E}_{jk} p_k = \theta_j$ where $p_k = \textrm{Tr}[\Pi_k \rho]$ are the probabilities for outcome $k$. The expected MSE for this estimator is \mbox{$\mathbb{E}[\textrm{MSE}] = \sum_{j,k} p_k(\mathcal{E}_{jk} - \theta_j)^2$}. For the state $\rho = \ket{0} \frac{1+r}{2} \bra{0} + \ket{1} \frac{1-r}{2} \bra{1}$, the expected MSE is minimized when $r_p = r$ and this saturates the N-H bound.

Without loss of generality, we have provided the analytical form of the POVM when the probe state is oriented along the $z$-axis. For any other state oriented differently, it is always possible to use a coordinate system that aligns its positive $z$-axis with the state and use the formulation given above before transforming the estimated state back to the original coordinate system. This change of basis doesn't affect the optimality of the measurement as the MSE based on trace norm is invariant under unitary transformations.

In principle, it is always possible that the estimated state will not lie within the Bloch sphere, i.e., $|\hat{\boldsymbol{\theta}}|>1$. However, in practical scenarios, as pointed out in Ref.~\cite{Petz2012EfficientMeasurements}, the probability of obtaining a physically impossible state will decay to zero with increasing repetitions of estimations due to the large deviation theory \cite{Varadhan2008LargeDeviations} as long as the estimator is unbiased. However, if the purity of the state is close to 1, the number of experiments required to guarantee the absence of a physically impossible estimate can be too big to realize experimentally. In this paper, we do not consider this issue because neither states with $r$ close to 1 nor an insufficient number of experiments occur.

\section{Methodology\label{sec:QCMeth}}

\begin{figure*}
    \centering
    \includegraphics[width = 0.8\linewidth]{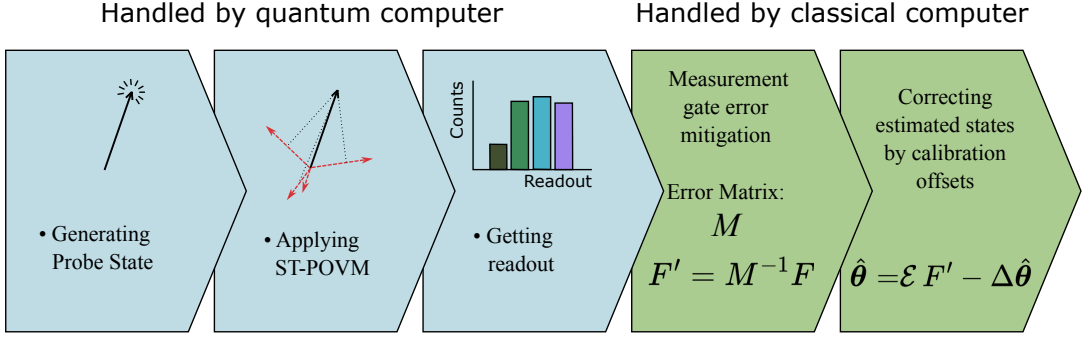}
    \caption{An illustration of the procedures involved in the demonstration. The first three stages are handled by the quantum computer, and the last two stages are handled by the classical computer. The classical operations take the frequencies of outcomes, $F$, and apply the measurement error mitigation. At the last step, by applying the pre-constructed estimator $\mathcal{E}$ and the calibration offset $\Delta\hat{\theta}$, we obtain the final estimates of the parameters}
    \label{fig:flowchart}
\end{figure*}

We implemented the ST-POVM on a superconducting quantum computer to demonstrate its feasibility and utility. An illustration of the general procedure of the demonstration is shown in Fig.\;\ref{fig:flowchart}. The demonstration involves both operations on quantum computers and classical computers. For the quantum operations, we generated the state and applied the ST-POVM before measuring it at the end. For the classical statistical operations, we utilize the measurement gate error mitigation technique \cite{Bravyi2020MitigatingExperiments} and a calibration procedure to improve our results. The procedure shown in Fig.\;\ref{fig:flowchart} performs estimation with a fixed number of probe states $N$. However, through this paper, we will subsample this data to obtain estimates of the quantum state based on different numbers of trials

In this section, we will first discuss the method of preparing probe states and then the method of implementing the POVM on the quantum computer. Finally, we will discuss the detailed implementation of the demonstration with error-mitigating techniques and calibration.

\begin{figure*}
    \centering
    \includegraphics[width = \linewidth]{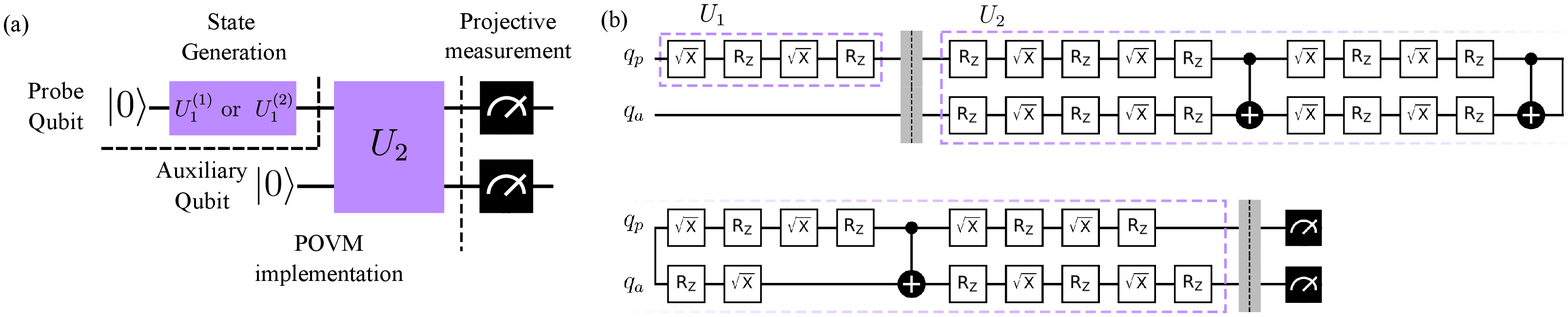}
    \caption{(a) A schematic diagram of the quantum circuit for the demonstration. $U_1^{(1)}$ or $U_1^{(2)}$ is a single qubit unitary used to prepare the appropriate probe state. $U_2$ implements the optimal ST-POVM. (b) An example of the circuit, all the circuits have the same structure. The dotted black lines separate the circuit into 3 steps, they correspond to $U_1$, $U_2$ and measurement gates in (a). R$_\textrm{Z}$ denotes the parametric $R_z$ gates with the corresponding input angles that realize $U_1$ and $U_2$.}
    \label{fig:Circuit}
\end{figure*}

\subsection{State preparation}
Quantum circuits only initialize pure states and the unitary quantum gates available do not change the purity of the initialized state in principle. Thus, to prepare a mixed state, we use the result that a mixed qubit state can be generated by a probabilistic combination of two pure states with probability $P_1$ and $P_2$. For a single qubit mixed state $\rho$, it is always possible to decompose the mixed state as follows:
\begin{equation}
    \rho = P_1 \ket{\phi_1}\bra{\phi_1} + P_2 \ket{\phi_2}\bra{\phi_2}\;.
    \label{eq:decomp}
\end{equation}
Thus, implementing an estimation scheme on a certain mixed state is equivalent to performing estimations with pure states randomly chosen between $\ket{\phi_1}$ and $\ket{\phi_2}$ with probabilities $P_1$ and $P_2$. The $U_1$ block in Fig.\;\ref{fig:Circuit}\;(a) generates the desired pure state $\ket{\phi_i}$ for the simulation of a mixed state.

$U_1$ is designed to bring the initial $\ket{0}$ state of the probe qubit to the pure state of either $\ket{\phi_1}$ or $\ket{\phi_2}$. The decomposition shown in Eq.\;\ref{eq:decomp} is not unique. In our demonstration, we use pure states $\rho(\boldsymbol{\theta})$ ($|\boldsymbol{\theta}| = 1$) that have the same $\theta_y$ and $\theta_z$ as the target mixed state $\rho$. The probabilities assigned to the two states will have a ratio of $P_1:P_2 = |\theta_x-\theta_x^{(2)}|:|\theta_x-\theta_x^{(1)}|$. $\theta_x$ is the $x$ axis projection of the mixed state. $\theta_x^{(1)}$ and $\theta_x^{(2)}$ are the $x$ axis projections of the two pure states $\rho_1(\boldsymbol{\theta}) = \ket{\phi_1}\bra{\phi_1}$ and $\rho_2(\boldsymbol{\theta}) = \ket{\phi_2} \bra{\phi_2}$. During the initialization of the mixed state $\rho$, a pure state will be randomly selected from $\{\rho_1, \rho_2\}$ with probabilities $P_1$ and $P_2$, respectively.

\subsection{Naimark Extension}
The quantum circuit uses two qubits for each instance of the chosen pure state. One is the probe qubit, and the other is the auxiliary qubit which helps the implementation of the POVM. The $U_2$ block in Fig.\;\ref{fig:Circuit}\;(a) represents the combined unitary operator of a series of two-qubit and single-qubit gates that entangle the probe qubit and the auxiliary qubit. This is to use the expanded Hilbert space for implementing the ST-POVM on the probe qubit.

We followed the principles of Naimark's dilation theorem \cite{Neumark,Gelfand1994OnSpace,Holevo2011ProbabilisticTheory}  to construct $U_2$, which requires an auxiliary qubit for our experiment. The $U_2$ constructed should entangle the probe qubit and the auxiliary qubit initialized in the $\ket{0}$ state in a particular way such that a chosen projective value measurement on the extended two-qubit state $E^+_i$ produces the same probability distribution as the original POVM, i.e.,
\begin{equation}
\label{eq:Naimark}
    \textrm{tr}(\rho\otimes \ket{0}\bra{0} U_2^\dagger E^+_i U_2) = \textrm{tr}(\rho \Pi_i) \;\;\;\; \textrm{for} \; \forall \rho \;.
\end{equation}
For more detail on Naimark's dilation theorem see Appendix \ref{apen:Naimark}.

In our demonstration, the quantum computer hardware supports projective measurements on each qubit, thus, the chosen measurement on the extended Hilbert space will be $E^+_i = \ket{e_i}\bra{e_i}$ where $\ket{e_i}$ denotes each of the four computational basis states: $\ket{00}$, $\ket{01}$, $\ket{10}$, $\ket{11}$. By Naimark's dilation theory and an numerical approach of implementing the POVM, we obtained the required $U_2$ that fulfils Eq.\;\ref{eq:Naimarkraw}. By applying $U_2$ on the two-qubit state and making projective measurements along the four computational basis states of the quantum computer available, we performed the ST-POVM on the probe qubit.

\subsection{Implementing on IBM quantum computer with autonomous package}

The decomposition of these unitary $U_1$ and $U_2$ are handled by the autonomous algorithms provided by IBM's quantum computing package, Qiskit \cite{Qiskit3}. An example of the native circuit running on one of the IBM quantum systems, \verb|ibmq_jakarta|, is shown in Fig.\;\ref{fig:Circuit}\;(b). We use this quantum system to generate all experimental results in this paper. For a single-qubit unitary, Refs.~\cite{Kitaev1997QuantumCorrection,Dawson2005TheAlgorithm} provide a scheme for the device with restricted sets of single-qubit operations to construct arbitrary single-qubit unitary gate. For a two-qubit unitary, Ref.~\cite{Vatan2004OptimalGates} shows that with 3 CNOT gates acting on the two qubits and some other single-qubit unitary operation, one can construct arbitrary two-qubit unitary operators.

\subsubsection{Readout noise and measurement gate error mitigation\label{sec:MGEM}}
Readout noise refers to the noise of the measurement gate. With up to 10\% error for assigning the $\ket{0}$ state as $\ket{1}$ or vice versa for some qubits on the IBM cloud computing devices, the readout noise posed a significant challenge to the accuracy of our repeated measurements. Thus, it is crucial to mitigate the readout noise produced by the measurement gates. 
In our demonstration, we followed the error mitigation scheme proposed in Ref.~\cite{Bravyi2020MitigatingExperiments} for minimizing readout noise. This error model proposes that the output frequencies vector $F'$ of a set of measurements have a linear correlation to the true frequencies $F$, which would be obtained on an ideal quantum computer. The linear correlation is as follows:
\begin{equation}
    F' = M F\;,
\end{equation}
where $M$ is a probability matrix that maps the ideal output $F$ to the experimental output $F'$. The method we implement first generates the error matrix $M$, a non-unitary operator describing the error. A set of 4 calibration circuits are run measuring only trivially generated states, $\ket{00}$, $\ket{01}$, $\ket{10}$, $\ket{11}$, which are assumed to be error-free. Then, we use the reverse correlation $M^{-1}$ to calculate what the true frequencies should be for the experimental runs:
\begin{equation}
    F = M^{-1} F'\;.
\end{equation}
Whenever we perform an offset calibration, we also generate the matrix $M$ so that the matrix $M$ is always up to date. This accounts for any possible drifting of hardware parameters.

\subsubsection{Time dependent calibration}
For any localized measurement that demands great precision, calibration before the measurement is recommended as it reduces systematic errors which cannot be reduced by repeated estimation, unlike random error due to statistical fluctuations. The execution of circuits on the quantum computer lasted only 12 hours. However, the entire execution is split into 4 time segments due to the availability of IBM's quantum computers. Thus, the possible time-dependent variables, such as the drift of hardware parameters and updated hardware calibrations may affect the raw output of the circuits. To account for this time-dependent issue, we performed calibrations regularly and compensated for the systematic offset for each corresponding main run. 

The calibration runs use the same $U_2$ (ST-POVM) but slightly different $U_1$ (state preparation) compared to the main runs. For precise calibration, we need to remove any possibility of circuit degeneration. The degeneracy of the circuit refers to the reduction of circuit complexity when initializing different states. For example, initializing a general state such as \mbox{$1/\sqrt{2}(\ket{0}+\mathrm{i}\ket{1})$} requires a more complex circuit than initializing state $\ket{0}$. To avoid degeneracy while keeping the control protocols on the quantum computers autonomous and robust, we adopted a semi-randomized approach to the initialization step of the main runs.

\begin{figure}
    \centering
    \includegraphics[width = \linewidth]{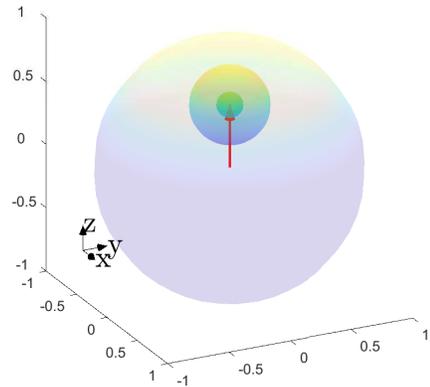}
    \caption{The ranges of the semi-randomized initialization scheme. The outermost sphere is the Bloch sphere, and the medium size sphere shows the range of randomly chosen calibration states. These calibration initial states will be used to find the offsets on parameter $\theta$. The smallest sphere shows the range of randomly chosen for the main runs. The medium sphere and the small sphere are centred around the intended true state shown by the vector arrow. (The sizes of spheres in this graph are exaggerated for improved visual clarity. The exact radii of these spheres are given in the main text)}
    \label{fig:Calibill}
\end{figure}

For every 2 of the main runs, we generate one calibration run consisting of 10 randomized mixed states in the unit sphere of radius 0.1 centred around the true state. We also randomized the exact mixed state in a small sphere of radius 0.01 to eliminate any degeneracy of state generation. The exact randomization scheme of the main run is as follows: We choose a mixed state for a set of 2 main runs randomly following a uniform distribution inside a small sphere of radius 0.01 centred around the mixed state we intended to simulate. Then, we initialize and measure the state with the exact procedure also implemented on the calibration circuits. The ranges of the semi-randomized initialization are shown in Fig.\;\ref{fig:Calibill}. The possible error generated from the semi-randomized approach will be discussed in the next section.

To correct the systematic error, we use the calibration runs to find the offsets for estimating each component of the Bloch vector. This is done by fitting the known input Bloch vector components to the estimated components via linear regression, with a fixed gradient of 1, to ensure that our estimates are locally unbiased. Next, we subtract the systematic offsets, $\Delta\bm{\hat{\theta}}$ in Fig.~\ref{fig:flowchart}, from the corresponding estimate of parameters obtained from the main runs. The method described above provided significant improvement to the overall accuracy, especially for estimates from large ensembles of identical copies of the qubit state.

\section{Results\label{sec:QCexp}}
In this section, we will present the results of the demonstration. We will also address our statistical method to analysed the raw results. 
For the very last stage of the demonstration shown as ``Correcting estimated states by calibration offsets" in Fig.\;\ref{fig:flowchart}, the input is the measurement results after being corrected by measurement gate error mitigation. This result is a series of random POVM outcomes chosen from the 4 possible outcomes $X_1,X_2,X_3$ and $X_4$ (such as \{$X_1,X_2,X_1,....,X_4,X_1$\}). The 4 possible outcomes correspond to the 4 POVM elements, and the length of the series is equal to the total number of probe states involved in the demonstration. In a practical setting, the number of probes may range from a few tens to millions. To comprehensively analyze the performance and the potential of the ST-POVM, we analyze the series of outcomes for the probe states by dividing them into groups of the same sizes, $N$.

From each group with $N$ outcomes, we calculate the frequency with which each POVM element is observed $F = (f_z,f_1,f_2,f_3)^\intercal$. Using Eq.\;\ref{eq:EstimateTheta}, we calculate the estimated value of $\boldsymbol{\theta}$, $\hat{\boldsymbol{\theta}}$, for each group. Then we calculate the MSE for each group and report the average MSE as the result. In this section, as we are working with multiple probes and the precision limit given by the \mbox{N-H} bound is MSE$\geq C_{\textrm{NH}}/N$, we expect the MSE to scale down with an increased number of probes (in each group) MSE$_N\rightarrow \textrm{MSE}_1/N$. In a practical setting, a large number of probes will be needed to obtain a satisfactory estimate of the state. Here we first present results using 100 probes to demonstrate the effectiveness of the ST-POVM for a lower number of probes. We then proceed to establish a model for the performance, in terms of MSE, when varying numbers of probes are used to estimate $\boldsymbol{\theta}$.

\subsection{MSE of 100-probe estimation}

\begin{figure}[ht]
    \centering
    \includegraphics[width = 1\linewidth]{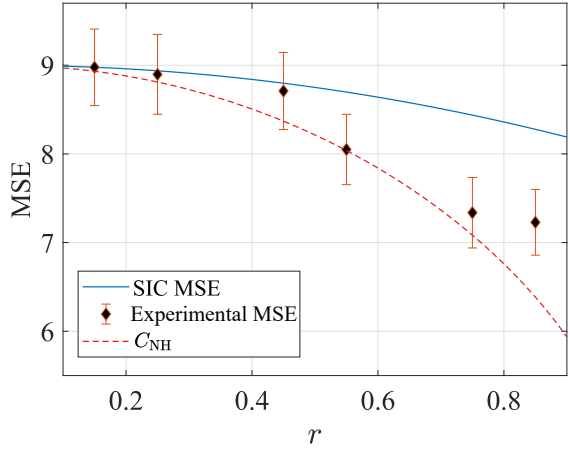}
    \caption{\label{fig:overall}The effective MSE using 100 probe states characterized by the length of the Bloch vector, $r$. The result is obtained from a total number of 180,000 identical probe states for each data point via repeated sub-sampling. The MSE is scaled to measure the effectiveness of the method per probe state, i.e., MSE = $N\;\textrm{MSE}_N$ where $N$ is the number of probes.
    The data points are the combined scaled MSE of the estimated state $\hat{\rho}(\hat{\theta}_x,\hat{\theta}_y,\hat{\theta}_z)$ when the group's size is 100. The blue (upper) line shows the theoretical MSE for a perfect SIC-POVM. The red dotted line indicates the N-H bound that lower bounds the MSE of any possible separable measurement. Error bars are one standard deviation obtained via bootstrapping.}
\end{figure}

We implemented the ST-POVM scheme on IBM's quantum computers. We obtained a set of results that shows the advantages and limitations of the measurement method we proposed in an experimental setting. The MSE in our demonstration conducted using 100 probe states is shown in Fig.\;\ref{fig:overall}. This result demonstrates that the POVM we suggest can saturate the N-H bound for some instances. Note, however, that in some cases, the average experimental MSE deviates from the theoretical prediction. We provide a brief discussion of this experimental error in the MSE obtained from sets of 100 probe states in this section, and a further elaborated analysis regarding the error in the experimental MSE using larger numbers of probes will be presented in Section.\;\ref{sec:Multi-probe}. 

\begin{figure}[ht]
    \includegraphics[width = 0.9\linewidth]{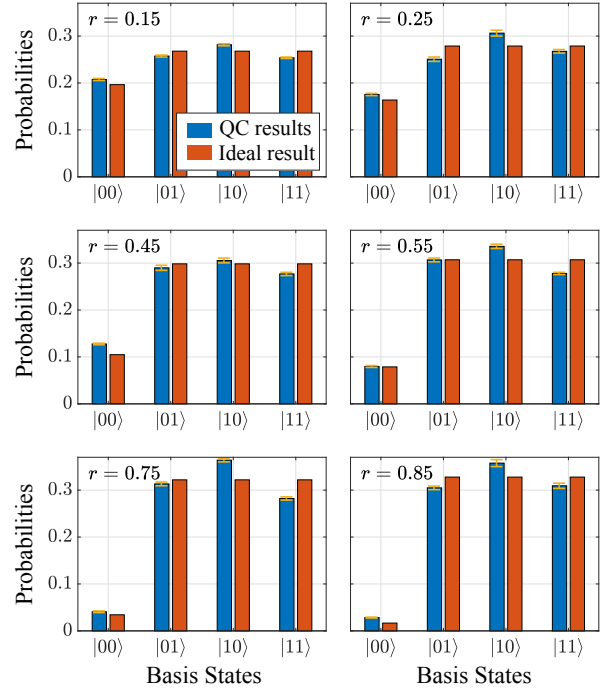}
    
    \caption{\label{fig:Dev_Ana} The deviation of the output state from the ideal state is demonstrated by the differences in probability for each measurement outcome. The probabilities of the outcomes marked by the 4 basis states are labelled by ``Quantum Computer (QC) result''. The probabilities of ideal outcomes are produced on simulated noiseless circuits and shown as the ``Ideal result''. The deviation of the amplitude from the ideal amplitude accounts for the time drift of the experimental setting across 6 sets of measurement. The error bar in the diagrams indicates the estimated statistical error of sampling from a multinomial distribution. Each set of measurements is performed within 30 minutes on IBM's quantum devices.}
\end{figure}

By analysing the entire series of POVM outcomes of 180,000 measurement for each $r$
, Fig.\;\ref{fig:Dev_Ana} hints at the kind of error we encounter when operating the quantum computer. For all 6 different quantum states, the probability amplitudes of the $\ket{10}$ state are consistently higher than the amplitudes of the ideal state, and the amplitudes of the $\ket{11}$ state are consistently lower than the ideal amplitudes. This could be the consequence of a systematic gate error or a gate alignment problem, but a definite conclusion about the origin of the error could not be made based on the data. As we have already performed measurement gate error mitigation before getting the result in Fig.\;\ref{fig:Dev_Ana}, we suspect that the influences of the measurement gate error on the data as a whole are relatively small compared to the observed deviation from the ideal result of up to 10\%.

We did not conduct any further investigation on the sources of errors with the POVM implementation as our focus is on the realization of the ST-POVM. To confirm and investigate the error contribution from the POVM implementation, we suggest a simpler measurement scheme such as the one shown in \cite{Huszar2012AdaptiveTomography}, which can potentially reconstruct any middle state in different stages of the circuit with a much lower error rate compared to the existing errors in the circuit.

\subsection{MSE with varying number of probe states\label{sec:Multi-probe}}
In the previous section, we presented the MSE analyzed in groups of 100 probe states. For practical use, we would like to consider various different numbers of probe states. Accounting for the possible error generated during the demonstration, we propose the following model for how the MSE scales with the number of probes used:
\begin{equation}
    \label{eq:MSE_model}
    \textrm{MSE}(N) = \frac{C}{N} + \delta\;,
\end{equation}
where $C$ is the experimental bound which depends on the specific implementation, and $\delta$ is a constant that accounts for the systematic error and drift of the experimental settings. As the N-H bound poses a fundamental limit to the state estimation problem, the $C$ value in the model above should not be less than the N-H bound, i.e., \mbox{$C\geq C_{\textrm{NH}}$}. In an ideal experiment, $C=C_{\textrm{NH}}$ and $\delta=0$.

Given a finite number of measurement outcomes ($N_{total}=180,000$ in our case), to accurately fit the model, we utilized repeated sub-sampling to get many more instances of the MSE$(N)$ than just $N_{total}/N$ sets. To conduct sub-sampling, we randomly divided the entire sample into $N_{total}/N$ sets and used the $N$ outcomes in each set to generate an instance of MSE$(N)$. Then we repeated the division of the entire data set multiple times to generate a greater number of separate instances of MSE$(N)$ values, from which an estimate of the average MSE$(N)$ will be calculated by averaging over all instances of MSE$(N)$. We observed no further convergence after reaching 10,000 instances of MSE$(N)$ for all probe states. Thus, we use 10,000 instances for all the results presented in this section.

\begin{figure}
    \centering
    \includegraphics[width = 0.9\linewidth]{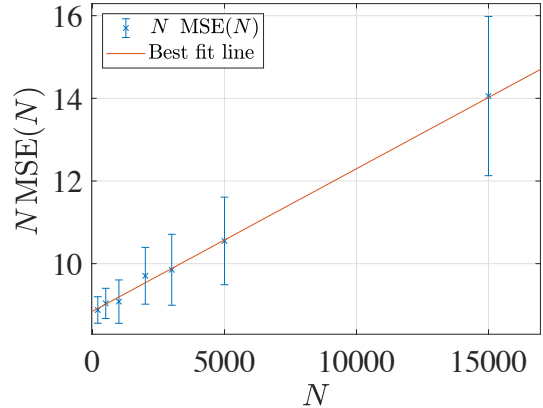}
    \caption{A demonstration of the model fitting according to Eq.\;\ref{eq:MSE_model1}. The figure is for $r=0.25$. The error bar shows the standard deviation of MSE for each number of probes obtained by bootstrapping.}
    \label{fig:MSE_model1}
\end{figure}

In Fig.\;\ref{fig:MSE_model1}, we fit a modified version of Eq.\;\ref{eq:MSE_model} shown below by linear regression
\begin{equation}
    \label{eq:MSE_model1}
    N\;\textrm{MSE}(N) = C + N\;\delta\;.
\end{equation}
With an increasing number of estimations, we can see that the specific MSE, $N\;$MSE$(N)$, increases almost linearly. This show that our model is a valid description of the results. In Table.\;\ref{tab:Decay_MSE_Fit}, for different probe states characterized by $r$, the values of the N-H bound, $C_{\textrm{NH}}$ and the values of the fitted parameters, $C$ and $\delta$ are shown. We attribute the deviation of $C$ from the N-H bound to the imperfect implementation of the POVM, and we attribute the value $\delta$ to imperfect calibration and drift of the experimental setup. In the following paragraphs, we will try to explain and validate the above claim.

\begin{table}[ht]
\caption{\label{tab:Decay_MSE_Fit}%
The result of the model fitting according to Eq.\;\ref{eq:MSE_model}.
}
\begin{ruledtabular}
\begin{tabular}{cccc}
\textrm{$r$}&
\textrm{$C_{\textrm{NH}}$}&
\textrm{$C$}&
\textrm{$\delta$}\\
\colrule
0.15 & 8.932 & $8.939 \pm 0.096$ & $2.72 \pm 0.92 \cross 10^{-5}$\\
0.25 & 8.810 & $8.937 \pm 0.066$ & $33.41 \pm 0.63 \cross 10^{-5}$\\
0.45 & 8.370 & $8.785 \pm 0.065$ & $7.45 \pm 0.62 \cross 10^{-5}$\\
0.55 & 8.038 & $8.108 \pm 0.115$ & $14.06 \pm 11.04\cross 10^{-5}$\\
0.75 & 7.083 & $7.358 \pm 0.072$ & $2.10 \pm 1.69 \cross 10^{-5}$\\
0.85 & 6.385 & $7.225 \pm 0.068$ & $13.07 \pm 0.65 \cross 10^{-5}$\\

\end{tabular}
\end{ruledtabular}
\end{table}

Due to the experimental imperfections, the POVM that is physically implemented by the imperfect circuit differs from the theoretical optimal POVM we proposed. As long as we treat all the quantum gates as unitary gates, the imperfect circuit always corresponds to some POVM acting on the first qubit. We can see in Fig.\;\ref{fig:Dev_Ana} that the probabilities of arriving at the 4 different POVM outcomes differ from the ideal probabilities calculated from the optimal POVM. Assuming that the initialization step gives negligible errors, this indicates the implementation of the POVM can be slightly different from the proposed POVM. This imperfect POVM is less optimal, and the variance resulting from the sub-optimal POVM will be greater than the N-H bound by an amount depending on the degree of error in the circuit.

We propose that the major contributor to the values of $\delta$ is the statistical error of the calibration runs. As we use a finite number of qubits for the calibration data, the calibration procedure will not be perfectly accurate. Thus, there is always a finite gap between the correction proposed by the calibration runs and the actual deviation between the means of the main measurement and the true values. Repeating the main runs will not eliminate this systematic error, so the effect of the slight inaccuracy of calibration will contribute to the constant term in Eq.\;\ref{eq:MSE_model}, $\delta$. We can verify this by calculating the expected standard deviation based on the finite amount of data used in calibration. For every 30,000 data points for the main measurement, we run 150,000 calibration measurements. The estimated standard deviation from the $n_{calib} = 150,000$ calibration runs will be approximately $C_{\textrm{NH}}/n_{calib} = $ $3\cross10^{-5}$ to $7\cross10^{-5}$, which is on the same scale as the $\delta$ listed in the Table.\;\ref{tab:Decay_MSE_Fit}. 

In addition to the statistical error of calibration, we propose other possible contributors to $\delta$ that are comparatively harder to estimate. One of them is random fluctuations of the experimental circuit (including the initialization and the POVM implementation) in terms of its average effect on an input state, i.e., fluctuation of the physical parameters when the circuit is implemented on the quantum computer. The fluctuation of measurement gate performance contributes to $\delta$ as well because the independent and identical performance of the measurement gate is one of the premises of the measurement error mitigation scheme discussed in Section.\;\ref{sec:MGEM}. Error reports based on the method of randomized benchmarking \cite{Magesan2011CharacterizingBenchmarking} can be fetched from the database of IBMQ. Still, the reports available based on the randomized method are insufficient for analysing the contribution to $\delta$ from the circuit. Other possible contributors can be the drift of experimental settings across different times and the statistical error in the result on the original 180,000 probes before sub-sampling.

\section{Complementary Simulations\label{sec:Comp_Sim}}
The method above and the N-H bound are designed to solve a localized problem, i.e., when the state is known relatively accurately, and we are trying to infer small changes in the state. However, in an alternative scenario, the qubit state can be completely unknown, or we only know approximately what the state might be. For such purposes, it is neither possible nor efficient to apply the ST-POVM measurement directly, as the orientation of the ST-POVM depends on the state. Thus, we present simulations of two adaptive methods in this section to demonstrate the potential usefulness of the ST-POVM in the scenario where we have some limited amount of prior knowledge. Adaptive approaches to tomography have been considered before both theoretically \cite{Huszar2012AdaptiveTomography} and experimentally \cite{Okamoto2017ExperimentalQubits,Kravtsov2013ExperimentalTomography,Mahler2013AdaptiveQuadratically}. In this section, we will first illustrate the simulations of an adaptive measurement scheme on states assumed to be completely unknown. Then we will consider optimal measurements for differing degrees of prior knowledge. We show that in the limit of a large number of probes, with an adaptive method utilizing the ST-POVM, the MSE approaches the N-H bound. We also show that as the amount of prior knowledge increases, the optimal measurement tends to the ST-POVM.

\subsection{Adaptive POVM using the SIC and ST-POVM\label{sec:adaptive}}

The simulation discussed in this section will show an adaptive scheme which gives an effective MSE that saturates the N-H bound asymptotically with the total probe number. We present a general guideline for implementing a two-step adaptive method, which uses the SIC-POVM initially before switching to the ST-POVM. We show that the optimal number of preliminary steps, given a fixed amount of total probe states, $N$, is proportional to $\sqrt{N}$ when $N$ is large. This section demonstrates the effectiveness of applying the ST-POVM on a more general problem of an unknown qubit state.

To measure a completely unknown state, one approach is an adaptive measurement scheme where the measurement of the next step will depend on the measurement result of the steps before it. In this way, the measurement applied for later steps will have reduced variance compared to the measurements applied before. In this example, we consider a two-step adaptive measurement where the SIC-POVM will serve as the preliminary measurement, and ST-POVM will be the secondary fine measurement. As the optimal ST-POVM is state-dependent, in the second stage, the ST-POVM chosen will be the most suitable measurement from the family of ST-POVMs based on the estimated state given by the preliminary step. After the two stages are completed, the final estimated state will be a weighted average of the result of the preliminary stage and the secondary stage. The final estimation of the parameters will be 
\begin{equation}
    \hat{\boldsymbol{\theta}} = W \hat{\boldsymbol{\theta}}^{(1)} + (1-W) \hat{\boldsymbol{\theta}}^{(2)}
    \label{eq:weighted estimator}
\end{equation}
where $0\leq W \leq1$ is the weight of the preliminary measurement. The superscript $(1)$ and $(2)$ denote the first step and the second step of the adaptive measurement. The MSE that we aim to minimize for optimizing this adaptive method is
\begin{equation}
    \textrm{MSE} = E[(\hat{\boldsymbol{\theta}}-\boldsymbol{\theta})^2]\;,
\end{equation}
where $\boldsymbol{\theta}$ is the probe state. With a given number of probes $N$ to complete the two-step adaptive measurement, we will need to allocate an appropriate amount of probes to the preliminary measurement and assign an appropriate value to the weight $W$ to minimize the MSE. 

\begin{figure}
\begin{tabular}{cc}
    \includegraphics[width=0.04\linewidth]{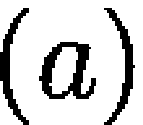} &
     \includegraphics[width=0.8\linewidth]{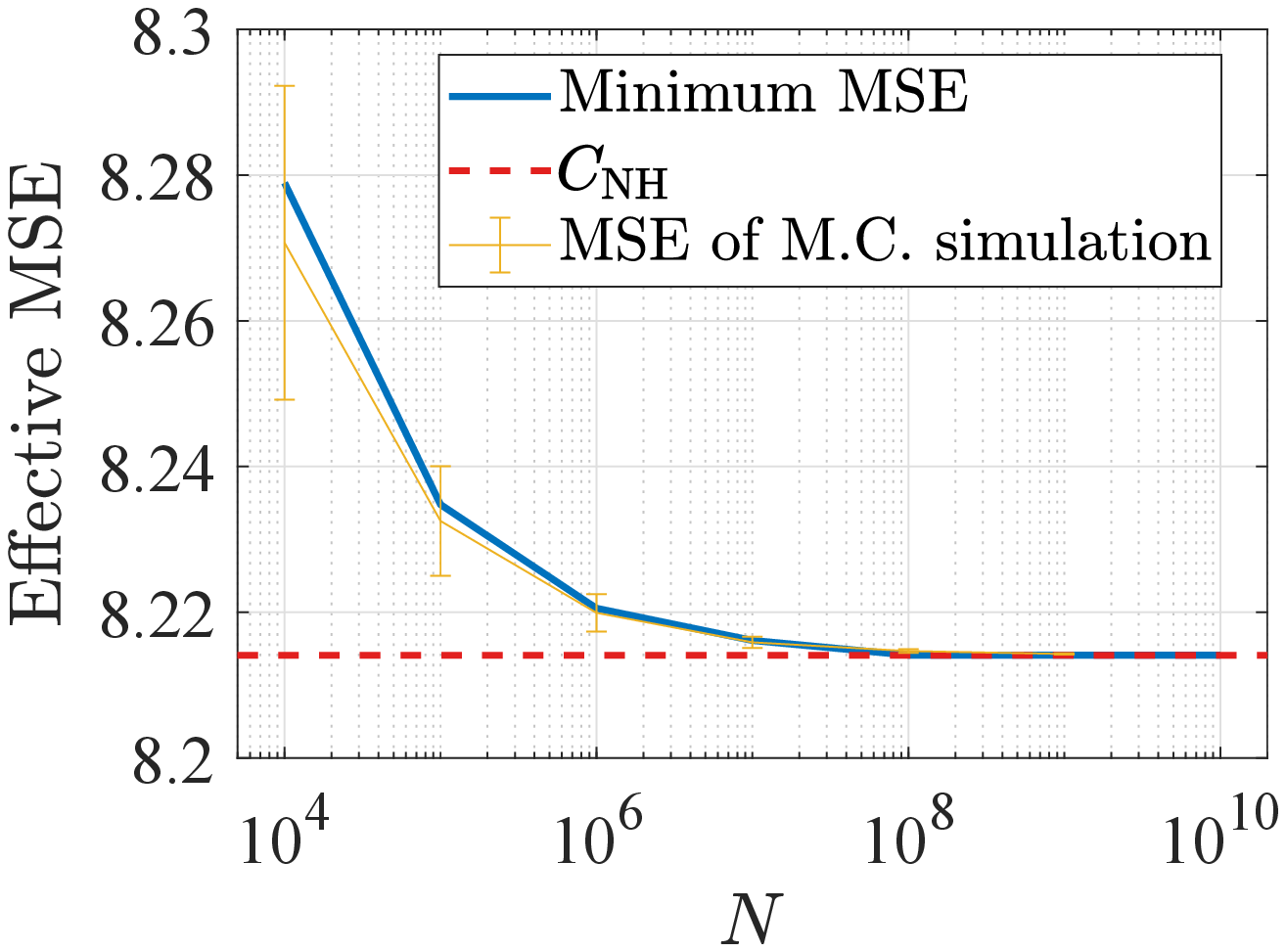}\\
     \includegraphics[width=0.04\linewidth]{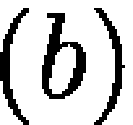}&
     \includegraphics[width=0.8\linewidth]{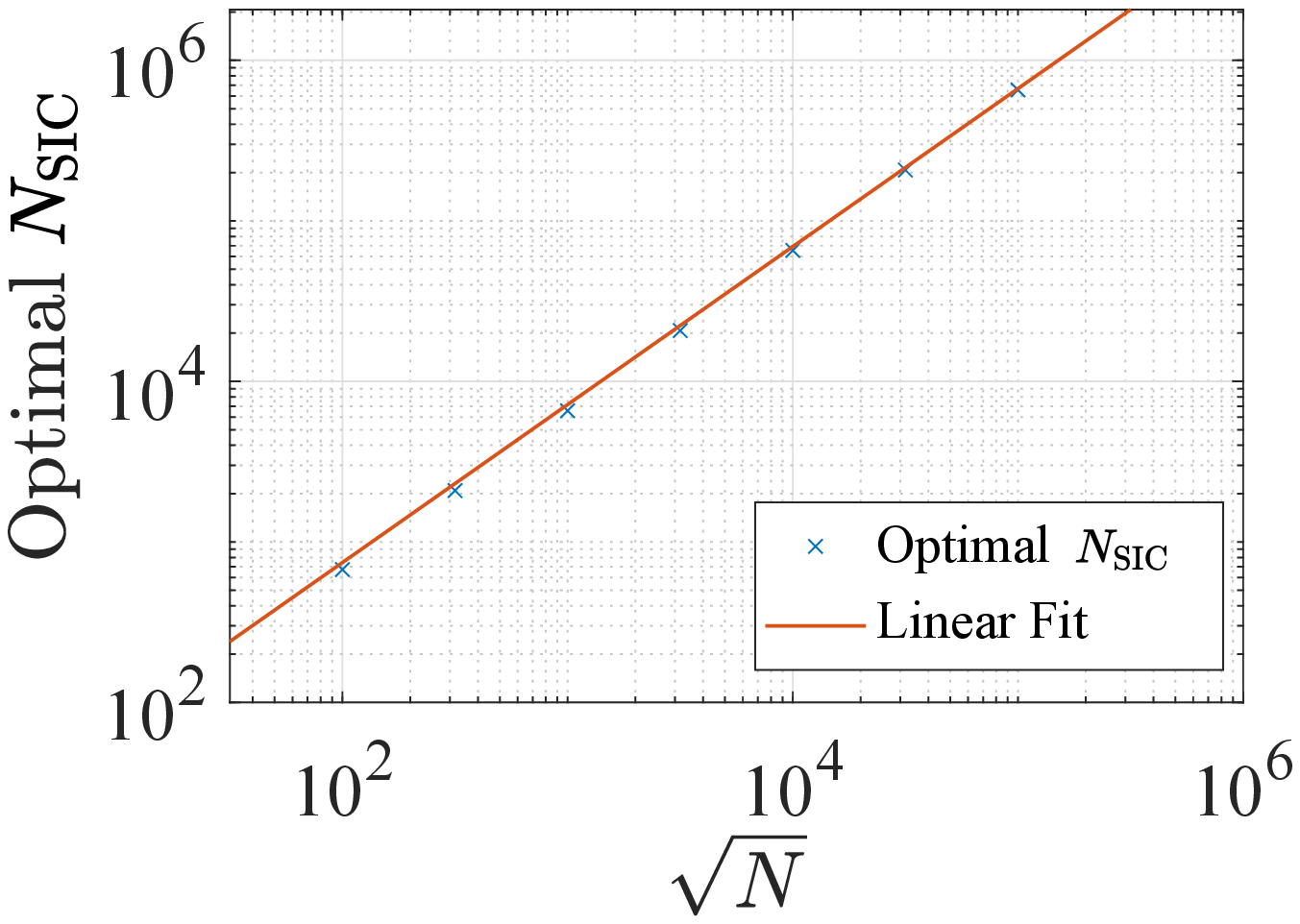}\\
\end{tabular}
\caption{\label{fig:MSEvsN}$(a)$ Asymptotic behaviour of the effective MSE when the number of total probes approaches infinity. A set of data generated from a Monte Carlo (M.C.) simulation is shown for reference. The error bar indicates the statistical error of the Monte Carlo simulation. The dotted red line shows the N-H bound. $(b)$ The change of the optimal preliminary step number, $N_{\textrm{SIC}}$ with different numbers of total probes, $N$. The linear fit verified Eq.\;\ref{eq:OptSIC}}
\end{figure}

To implement simulations with up to $10^{10}$ total available probes, we choose to use an approximated analytical approach instead of Monte Carlo sampling. We will justify the approximation used in this approach below and, in Fig.\ref{fig:MSEvsN} ($a$), a referential set of data is given to further demonstrate the asymptotic accuracy of the approximated analytical approach. For this analytical approach, the central limit theorem that applies to a large number of repeated sampling facilitates our analysis. For a large number of available probe states, by the multidimensional central limit theorem\cite{Vaart1998AsymptoticStatistics}, the distribution of the estimated state for each stage can be well approximated by a multivariate normal distribution centred around the true state with covariance matrix $\Sigma_N = \Sigma/N$. However, as shown in Fig.~\ref{fig:MSEvsN} ($a$), the approximation tends to introduce more error in the low probe number regime and the results obtained converge to the Monte Carlo Simulation in higher probe number.

The central limit theorem also leads to a simplification of finding the optimal weight $W$ from Eq.\;\ref{eq:weighted estimator}. Assuming that the results obtained from both stages are unbiased, the estimated states for the two stages are uncorrelated on all three parameters $\theta_x$, $\theta_y$ and $\theta_z$. For $\theta_j$, this translates to 
\begin{equation}
    Cov(\theta_j^{(1)},\theta_j^{(2)}) = 0\quad (j=x,y,z)\;, 
    \label{eq:uncorrelated}
\end{equation}
where $Cov(\;\cdot\; ,\;\cdot\;)$ is the covariance between the two random variables. Note that the two outcomes from the two stages remain statistically dependent on each other as the parameter of the second stage depends on the result of the first stage. Let the MSE of the preliminary stage be MSE$_1$, and the MSE of the secondary stage be MSE$_2$. By Eq.\;\ref{eq:uncorrelated}, we simplify the total MSE based on Eq.\;\ref{eq:weighted estimator} to be
\begin{equation}
\label{eq:MSEsimp}
    \textrm{MSE} = W^2 \textrm{MSE}_1 + (1-W)^2 \textrm{MSE}_2\;.
\end{equation}
The MSE is minimized by setting
\begin{equation}
\label{eq:Woptimal}
    W = \frac{\textrm{MSE}_2}{\textrm{MSE}_1+\textrm{MSE}_2}
\end{equation}
Using the optimal value of the weight $W$, we can find the average MSE for different numbers of allocated preliminary measurements by Eq.\;\ref{eq:weighted estimator}. For a given preliminary outcome $\hat{\boldsymbol{\theta}}^{(1)}$, we have MSE$_1=(\hat{\boldsymbol{\theta}}^{(1)}-\boldsymbol{\theta})^2$. If the function MSE$_2(\hat{\theta}^{(1)})$ is known, given initial parameter, $\theta$, we are able to evaluate both MSE$_1$ and MSE$_2(\hat{\theta}^{(1)})$. By Eq.\;\ref{eq:MSEsimp} and Eq.\;\ref{eq:Woptimal}, a MSE value that depends on the preliminary estimate $\hat{\theta}^{(1)}$ can be constructed. The method that we use to find the function, MSE$_2(\hat{\theta}^{(1)})$, is a change of the coordinate system. Based on the preliminary estimate, $\hat{\theta}^{(1)}$, we switch to a coordinate system such that the predicted state lies along the negative $z$-axis. This means that the optimal POVM, used in the second stage, is oriented along the positive $z$-axis. The MSE obtained in this setting is equal to the MSE in the original coordinate system because the MSE derived from the trace norm between states is invariant under a unitary rotation. Then, after transforming $\hat{\theta}^{(1)}$ back to the original coordinate system, we have the function MSE$_2(\hat{\theta}^{(1)})$.

Lastly, we will need to iterate through different numbers of probes allocated for the two stages to find the optimal number of preliminary measurements together with the minimum MSE achieved by the two-step adaptive method. Thus, we perform numerical integration with a fixed number of preliminary measurements using the SIC-POVM, $N_{SIC}$, as shown below:
\begin{equation}
\label{eq:analytic_ST}
    \overline{\textrm{MSE}} = \iiint_S \textrm{MSE}(\hat{\boldsymbol{\theta}}^{(1)}) \; \text{pdf}(\hat{\boldsymbol{\theta}}^{(1)}) \;\,d\hat{\boldsymbol{\theta}}^{(1)}\;,
\end{equation}
where pdf$(\hat{\boldsymbol{\theta}}^{(1)})$ is the probability density function of the multivariate normal distribution that we use to approximate the discrete distribution of the preliminary outcomes. $S$ is the entire parameter space with $|\hat{\boldsymbol{\theta}}^{(1)}|<1$.

\begin{figure}[ht]
    \centering
    \includegraphics[width = 0.9\linewidth]{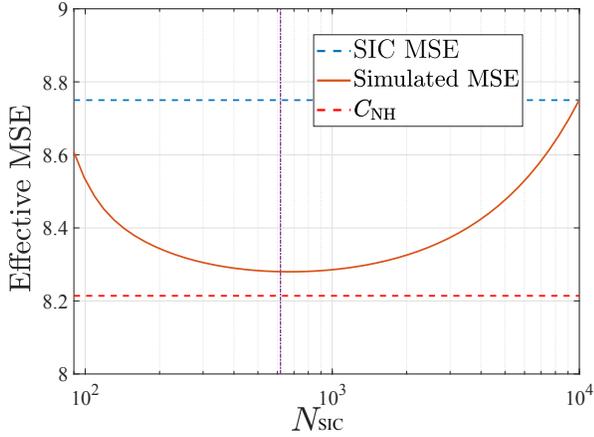}
    \caption{The effective MSE given 10,000 probe states, $N\;\textrm{MSE}(N)$, for different number of allocated preliminary steps, $N_{\textrm{SIC}}$. There is an optimal number of preliminary step $N'_{\textrm{SIC}}$ that gives a minimum MSE. The MSE obtained when only SIC-POVM is used is indicated by the upper flat line. The lower flat line indicates the N-H bound. The optimal $N_{\textrm{SIC}}$ is indicated by the vertical line.} 
    \label{fig:MSEvsNsic}
\end{figure}

We provide the detailed graphed result for one sample qubit state in Fig.\;\ref{fig:MSEvsNsic} and Fig.\;\ref{fig:MSEvsN}, and the results of 5 other states are presented in Table.\;\ref{tab:Clist} For the detailed examples, we use the state $\rho = (I + \Vec{\sigma}\cdot\boldsymbol{\theta})/2$ where $\boldsymbol{\theta} = (0,0,0.5)^\top$ is the vector of parameters. For this specific state, Fig.\;\ref{fig:MSEvsNsic} shows the change in MSE when allocating different numbers of qubits to the SIC-POVM preliminary experiments when the total number of qubits is 10,000. It shows that there is one and only one minimum MSE. Either more preliminary steps or fewer preliminary steps will increase the MSE. For 10,000 qubits, the optimal number of preliminary SIC-POVM experiments is $673$ (without accounting for modelling error). In this case, the effective MSE is only 0.8\% above the N-H bound at 8.28. In contrast, for a pure SIC-POVM approach, the effective MSE will be more than 6.5\% above the N-H bound at 8.75. For a larger number of total probes, the effective MSE is closer to the N-H bound and will eventually converge to the N-H bound when $N\rightarrow \infty$. This asymptotic behaviour is shown in Fig.\;\ref{fig:MSEvsN} (a). We also found that for this particular two-step adaptive measurement on the state specified above, the optimal number of preliminary experiments, which we denote $N'_{\textrm{SIC}}$,  can be approximated by 
\begin{equation}
    N'_{\textrm{SIC}} = B \sqrt{N}\;,
    \label{eq:OptSIC}
\end{equation}
where B is a constant, which depends on the state's purity and orientation. Fig.\;\ref{fig:MSEvsN} (b) shows the change of $N'_{\textrm{SIC}}$. By linear regression, we found that the $B = 6.55$ is the coefficient.

\begin{table}[h!]
\caption{\label{tab:Clist}%
The coefficient $B$ of various states on the $z$-axis. The value of $B$ grows to infinity when the $\theta_z$ value approaches 0, 1, and -1. Thus, we did not give a value of $B$ for $\theta_z = 0$. The algorithms we have designed can solve the problem of finding the optimal number of preliminary steps with arbitrary parameters. This table serves as a general guideline of how many preliminary numbers are preferred and also as a demonstration of our algorithms. 
}
\begin{ruledtabular}
\begin{tabular}{c|ccccccc}
$\theta_z$ & -0.9 & -0.5 & -0.3 & 0 & 0.3 & 0.5 & 0.9 \\
$B$ & 8.39 & 6.76 & 10.35 & NA & 10.28 & 6.55 & 5.24

\end{tabular}
\end{ruledtabular}
\end{table}

We have included the suggested coefficient $B$ for other states when the ST-POVM is aligned along the $z$-axis in Table.\;\ref{tab:Clist}. The initial SIC-POVM in our simulated experiment always points downward, similar to the `squashed' tetrahedron shown in Fig.\;\ref{fig: STH_vis}. We also found that the value of $B$ grows when the $\theta_z$ value approaches 0, 1, and -1. For 1 and -1, as the ST-POVM becomes more specialized, it requires more preliminary steps to ensure a correct input estimated state until at $r=1$, the formulation introduced in Section.\;\ref{sec:OptPOVM} breaks as the bottom left entry, $b+r$, in the estimator matrix $\mathcal{E}$ in Eq.\;\ref{eq:EstimateTheta} becomes infinity. This incompatibility of the ST-POVM with pure states is expected because setting the preliminary expectation of $r=1$ eliminates the uncertainty of estimating $\hat{\theta}$ in the radial direction. For example, the state $\ket{0}$ has a vanishing $z$ direction variance:
\begin{equation}
    \mathbb{E}[(\hat{\theta_z} - \theta_z)^2] = \sum_{i=0,1} p_{\ket{i}} (\theta_{z,\ket{i}}-1)^2 = 0
\end{equation}
This changes the three-parameter estimation problem to a two-parameter problem. Thus, setting $r=1$ violates the assumption of estimating 3 parameters simultaneously for the ST-POVM estimation scheme. For $\theta_z= 0$ the SIC-POVM is already the optimal POVM. Thus, any further adaptive measurement will be less optimal.

Note that Table.\;\ref{tab:Clist} is computed assuming that the state is known. In an experiment with a completely unknown probe state, it will not be possible to use the optimal $B$ value or the optimal weight in Eq\;\ref{eq:Woptimal}. For an experiment with a flexible amount of probe states, one could use approximately $20 \sqrt{N}$ of the probes for the preliminary SIC-POVM without inducing significant overhead for the ST-POVM while maintaining high suitability for the follow-up measurement. In this simulation, the relative orientation between the initial state and the SIC-POVM will change the coefficient $B$. The reason for this is that although estimations given by the SIC-POVM give a uniform MSE$ = 9- r_p^2$ (Eq.\;\ref{eq:MSE_SIC}) regardless of the orientation of the POVM, the distribution of SIC-POVM estimates is not invariant with a unitary rotation. This can be explained by the discrete structure of the SIC-POVM. We should bear in mind that an arbitrary rotation of the tetrahedron representing the SIC-POVM does not guarantee the orientation invariance like a rotation acting on the sphere does. This is what gives rise to the asymmetry in Table.\;\ref{tab:Clist}.

\subsection{Evolving ST-POVM for general adaptive measurement}
\begin{figure*}
\includegraphics[width=0.95\textwidth]{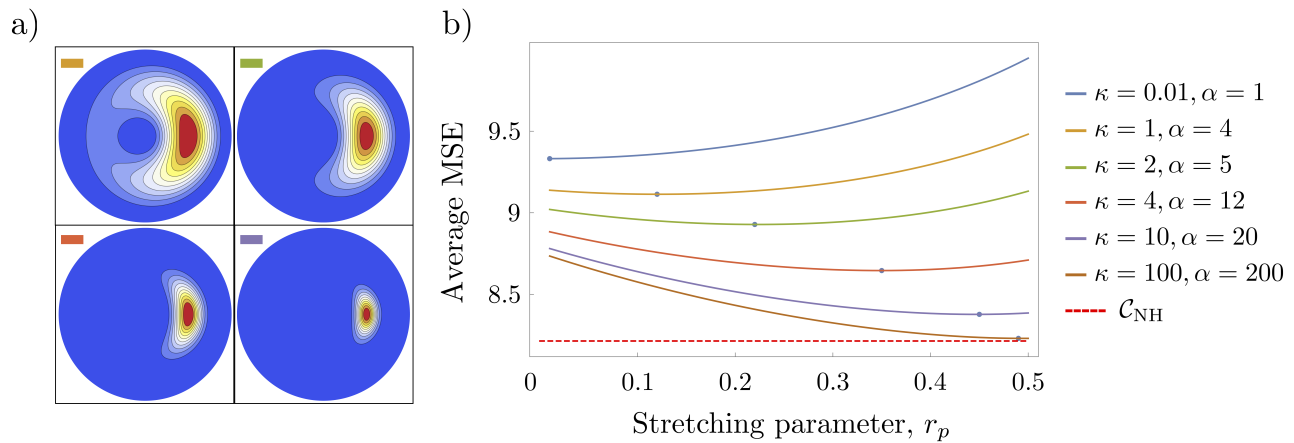}
\caption{\textbf{Minimizing the average MSE given a certain prior distribution of states. } (a) Examples of the prior distribution of states for a cross-section of the Bloch sphere in the $x-y$ plane. Legend is the same as in (b). (b) Average MSE as a function of the stretching parameter, $r_p$, for different prior distributions. The blue dots indicate the minimum MSE for each prior distribution. As the distribution gets more sharply peaked, the optimal $r_p$ tends to the $r$, which saturates the N-H bound, and the MSE tends to the limit set by the N-H bound.}
\label{fig:bayesrisk}
\end{figure*}

In the previous simulation, we considered a simple two-step adaptive measurement. This MSE approached the quantum limit, the N-H bound, asymptotically when the number of available probe states tends to infinity. Thus, there is little space for other more sophisticated adaptive schemes to improve upon. However, when the total number of probes is low, the evolving adaptive measurement utilizing the Bayesian estimation\cite{ChristianP.Robert2007TheChoice} procedure is more advantageous. During an evolving adaptive measurement, the POVM implemented is adjusted more frequently according to the results obtained from the preceding steps. In such an adaptive measurement scheme, the stretching parameter\footnote{The stretching parameter, $r_p$, may be seen as a representation of the degree to which the squashed tetrahedron POVM is squashed. When $r_p = 0$, there is zero squashing and the ST-POVM converges to the SIC-POVM}, $r_p$, (first shown in Section.\;\ref{sec:OptPOVM}) is updated over time so that, at each stage, the optimal POVM is being used.

In each iteration of an evolving adaptive measurement, a fixed algorithm will decide the most suitable POVM based on the prior distribution of states given from the previous measurement results. The objective of the entire adaptive procedure will be to improve the sharpness of the posterior distribution so that the uncertainty of the estimated state is minimized. After a POVM is decided, it will be applied to one of the probes. Using the results of this POVM and following the formulation of Bayesian estimation, one could obtain the posterior distribution, which will serve as the prior for the next iteration. In this section, we will give a paradigmatic example of finding the most suitable POVM for a given Bayesian prior distribution of state based on the minimization of Bayesian risk \cite{Rubio2019QuantumData}. We emphasise that this is different to simulating the full adaptive measurement scheme. This section merely shows what is the optimal POVM given a certain prior distribution.

For the purpose of this demonstration, we assume that the prior distribution is invariant under a rotation around its radial axis. Thus, the state distribution under the spherical coordinate system, $f(r,\lambda, \phi)$, can be decomposed to a separable form, 
\begin{equation}
    \label{eq:f}
    f(r,\lambda,\phi) = f_r(r) f_{\lambda\phi}(\lambda,\phi)\;.
\end{equation}
For simplicity, we use the 3-D Von--Mises Fisher distribution \cite{Fisher1953DispersionSphere} for the probability density function $f_{\lambda\phi}$ with a unified concentration parameter $\kappa$ for all values of $r$. For the distribution along the radial direction, $f_r$, we use the Beta distribution with two positive shape parameters $\alpha$ and $\beta$. We set $\beta=\alpha(1-r)/r$ where $r$ is the distance from the centre of the distribution to the origin of the Bloch sphere.

For a given prior distribution, we will assume that we always point our ST-POVM along the direction of the distribution's radial axis. The Bayesian risk, $\bar{\epsilon}(r_p)$, when using a POVM with parameter $r_p$ can be calculated as follows~\cite{Rubio2019QuantumData}:
\begin{equation}
\bar{\epsilon}(r_p)=\int dr \,d\lambda \, d\phi\,\text{MSE}(\boldsymbol{\theta}_p,r,\lambda,\phi) f(r,\lambda,\phi)\;,
\end{equation}
where $\text{MSE}(\boldsymbol{\theta}_p,r,\lambda,\phi)$ is the MSE of a measurement on the probe state $\rho(r,\lambda,\phi)$, where $(r,\lambda,\phi)$ are the spherical coordinates of the probe state in the Bloch sphere, when the ST-POVM is orientated along the direction of $\boldsymbol{\theta}_p$ with parameter $r_p = |\boldsymbol{\theta}_p|$. Due to the symmetry of the prior distribution, we choose the orientation of the POVM to be the same as the orientation of the distribution. Thus, the chosen ST-POVM and the distribution are invariant when rotating around the axis of $\boldsymbol{\theta}_p$. Consequentially, without loss of generality, we write the Bayesian risk on the left-hand side as a function of $r_p$. The equation above indicates that the Bayesian risk, $\bar{\epsilon}(r_p)$, is the average of the MSE we first defined in the introduction over the distribution of quantum states using the same POVM with parameter $r_p$.

We compute the Bayesian risk for a range of distributions in the form of Eq.\;\ref{eq:f}, centred on a state with $r=0.5$. Examples of this prior distribution are shown in Fig.~\ref{fig:bayesrisk} (a). For each distribution, to find the best ST-POVM with parameter $r_p$, the Bayesian risk is calculated for a range of different stretching parameters.

The results of the calculation are shown in Fig.~\ref{fig:bayesrisk} (b). It shows that with an increase of the parameter $\kappa$ and $\alpha$, the minimum Bayesian risk approaches the N-H bound. As an increase in $\kappa$ and $\alpha$ implies a more localized prior distribution, the results also suggest that the ST-POVM is applicable and asymptotically optimal in an evolving adaptive measurement scheme that utilizes Bayesian estimation. It is also worth noting that the ST-POVM should be close to the optimal measurement in terms of MSE for the later stages (when the uncertainties of the priors are low) of any evolving adaptive tomography scheme on the single qubit state. As the optimal measurement with zero prior knowledge was found to be the ST-POVM with a stretching parameter $r_p=0$, i.e. a SIC-POVM, this framework offers a nice way to visualize the evolution of the optimal measurement.

\section{Conclusion}
In this paper, we have mapped qubit tomography to a quantum multiparameter estimation problem. Based on the multiparameter estimation bound for separable measurements (N-H bound), we found the theoretically optimal measurement for the local tomography problem which we call the ST-POVM. As such local measurements that characterize the qubit state have direct implications for quantum computing, we performed a demonstration on an IBM superconducting quantum computer, the result shows the saturation of the N-H bound. In Section.\;\ref{sec:Comp_Sim}, we argued that the ST-POVM could also be effective in the non-localized setting by showing the results of two complementary simulations of two different types of adaptive measurement.

In this paper, single-qubit state estimation is the focus. However, to better evaluate the performances of a more general quantum circuit, one needs to consider more than one qubit. We will briefly discuss two possible ways of extending the scope of the demonstration. Previous studies have shown that entangling measurements help reduce the experimental errors \cite{Hou2018DeterministicWalks,Conlon2023ApproachingPlatforms,Roccia2017EntanglingQubits} and can be useful in state discrimination~\cite{Conlon2023DiscriminatingQubit}. Thus, the first extension can be to evaluate more than one copy of the qubit state at once, as the entanglement between multiple probes state will be able to reduce the experimental errors further. While the exact method that attains the N-H bound for multiple copies of a state $\rho^{\otimes n}$ remains unknown, the ST-POVM with its corresponding estimator introduced in this paper may have implications in finding a series of POVMs exploiting the symmetry of a multi-copy quantum system that saturates the N-H bound. For the SIC-POVM, a two copies extension has been tested on the photonic system \cite{Hou2018DeterministicWalks}, which suggests a possible generalization of the ST-POVM to multiple copies of a qubit system.

On the other hand, a less complicated extension of the results in this research would be to reduce the complexity by considering two of the three parameters. The search for the two-parameter optimal POVM is promising because it has been shown that there is an equivalent of the SIC-POVM in this problem, the conditional SIC-POVM \cite{Petz2014ConditionalSIC-POVMs,Ohno2015ExamplesSIC-POVMs}. It is optimal when the figure of merit is the average Hilbert--Schmidt norm deviation. Alternatively, for a general Hilbert space of dimension $d$ with $d(d+1)/2$ free parameters, there might be a conditional optimal POVM with a similar structure as the ST-POVM when 3 of the $d(d+1)/2$ parameters are unknown, similar to the 3 parameters for the qubit tomography problem, $\theta_x$, $\theta_y$ and $\theta_z$.

In summary, the result of our demonstration validates the theory and illustrates the power of a general quantum computer in implementing POVM measurements. The scheme we developed may also provide an opportunity for understanding the characteristics of the qubits and the quantum gates by facilitating a more well-rounded assessment of any single-qubit output state.

\begin{acknowledgments}
 We wish to acknowledge Prof Jun Suzuki for his valuable and unreserved advice, Dr Yong-Su Kim for his insightful questions, and Chenyue Gu for valuable proofreading.

 We acknowledge the use of IBM Quantum services for this work. The views expressed are those of the authors and do not reflect the official policy or position of IBM or the IBM Quantum team.
\end{acknowledgments}

\appendix

\section{Naimark's dilation theorem }
\label{apen:Naimark}
Given any POVM $\{\Pi_i\}$, Naimark's dilation theorem \cite{Neumark,Gelfand1994OnSpace,Holevo2011ProbabilisticTheory} guarantees the existence of an isometry $\tilde{V}\;:\; \mathcal{H} \rightarrow \mathcal{H^+}$, and
a projective measurement $\{E_i^+\}$ on an extended Hilbert space $\mathcal{H}^+$ (with the same number of elements), which satisfy  
\begin{equation}
\label{eq:NaimarkTheory}
    \tilde{V}^\dagger E_i^+ \tilde{V} \rho = \Pi_i \rho
    \;\;\;\; \textrm{for} \; \forall \rho \;.
\end{equation}
As we shall see later, the above equation is solvable for $\mathcal{H^+}$ being a two-qubit Hilbert space. In the following discussion, we will use a single auxiliary qubit to realize the POVM. Because the auxiliary qubit on a quantum computer is initialized at the $\ket{0_a}$ state, we consider an isometry $\tilde{V}$ with the following effect on a mixed state $\rho$:
\begin{equation}
    \tilde{V} \rho = U_2 (\rho \otimes \ket{0_a}\bra{0_a}) 
\end{equation}
This simplifies Eq.\;\ref{eq:NaimarkTheory} to
\begin{equation}
\label{eq:Naimarkraw}
    \bra{0_a}U_2^\dagger E_i^+ U_2\ket{0_a} \rho = \Pi_i \rho
    \;\;\;\; \textrm{for} \; \forall \rho \;.
\end{equation}
For projectors $E_i^+$, the unitary $U_2$ effectively rotates the hyperplane that the operator projects onto. Thus, we can write the operators on the extended Hilbert space $U_2^\dagger E_i^+ U_2$ as another projector, $\ket{\psi^+_i}\bra{\psi^+_i}$. Now, we force the operators acting on $\rho$ on each side to be the same, and we have:
\begin{equation}
\label{eq:psi+}
    \bra{0_a}\ket{\psi^+_i} = \sqrt{r_i} \ket{\psi_i}\;,
\end{equation}
where $r_i$ and $\ket{\psi_i}$ defines the ST-POVM in Sec.\;\ref{sec:OptPOVM}. A numerical calculation can then be performed to find all four elements of orthonormal $\{\ket{\psi^+_i}\}$ given Eq.\ref{eq:psi+}. It is also worth noting that Eq.\;\ref{eq:Naimarkraw} also guarantees that the measurement outcome distribution is unchanged by this process, which is crucial for the success of the estimation, i.e.,
\begin{equation}
\label{eq:Naimark1}
    \textrm{tr}(\rho\otimes \ket{0}\bra{0} U_2^\dagger E^+_i U_2) = \textrm{tr}(\rho \Pi_i) \;\;\;\; \textrm{for} \; \forall \rho \;.
\end{equation}



%

\end{document}